\newcommand{\acs}{}
\newcommand{\gls}{}
\newcommand{\acp}{}

\newcommand{\glsplural}{}
\newcommand{\g}{$\gamma$}

\documentclass{PoS}
\RequirePackage{nccmath}
\RequirePackage{enumerate} 
\RequirePackage{subcaption}
\RequirePackage{wrapfig} 

\title{The Topo-trigger: A new stereo trigger for lowering the energy threshold of IACTs }

\ShortTitle{Topo-trigger}

\author{\speaker{Rub\'en L\'opez-Coto}\\
        IFAE, Campus UAB, E-08193 Bellaterra, Spain\\
	E-mail: \email{rlopez@ifae.es}
        }

\author{Daniel Mazin\\
Institute for Cosmic Ray Research, University of Tokyo, 277-8582 Chiba, Japan\\
}        

\author{Riccardo Paoletti\\
Universit\`a  di Siena, and INFN Pisa, I-53100 Siena, Italy\\
}             

\author{Oscar Blanch, Juan Cortina\\
        IFAE, Campus UAB, E-08193 Bellaterra, Spain\\
        }

\abstract{The purpose of the hardware presented in this contribution is to decrease the energy threshold of the MAGIC telescopes without significantly increasing the data acquisition rate. To achieve this purpose, we developed an additional level of trigger that relies on the location in both MAGIC cameras where the trigger is issued to rule out accidental events. This allows to decrease the Discriminator Threshold (\acs{DT}), which results in a reduction of the energy threshold of the instrument. We simulated the Topo-trigger concept using the standard MAGIC Monte Carlo (MC) and tested it with real telescope data. In this paper we show the concept and results of these tests.

}

\FullConference{The 34th International Cosmic Ray Conference,\\
		30 July- 6 August, 2015\\
		The Hague, The Netherlands}

\begin{document}


\section{Limitations of the trigger system in the MAGIC telescope}
\label{sec:limitations_trigger_magic}
The trigger system in the MAGIC telescope is hardware limited at several stages. Every time a trigger is issued, the  trigger system is busy for 100 ns, not accepting any other trigger in this time. Regarding the stereo trigger, the maximum stereo rate the current \gls{DAQ} can record is $\sim$3 kHz \cite{DAQ_Diego}.

\begin{minipage}[h]{0.57\textwidth}
  \includegraphics[width=\textwidth]{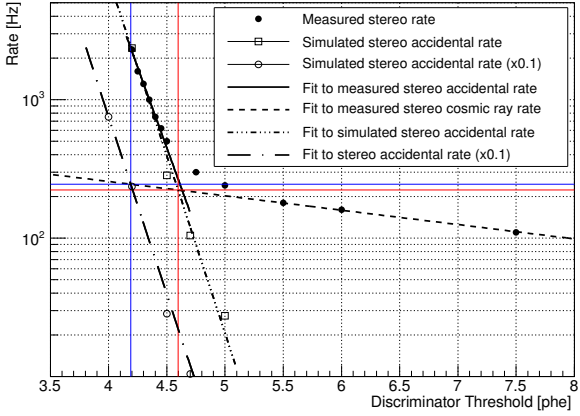}
\captionof{figure}{Measured and simulated stereo trigger rate for the MAGIC telescopes. The crossing point of the red lines determines the current operation point in the MAGIC telescope. The crossing point of the blue lines marks the operation point where we would go by decreasing the accidental rate to 10\% of its value and maintaining the same rate recorded.}
\label{trigger_rate_measured}
\vspace{0.5 cm}
\end{minipage}
 \hspace{0.01\textwidth}
\begin{minipage}[h]{0.35\textwidth}

The simulated stereo accidental trigger rate (open squares) and measured stereo accidental trigger rate (filled circles) of MAGIC are shown in Figure \ref{trigger_rate_measured}. We are currently working at the crossing point between the extrapolation of the stereo cosmic ray trigger rate and the stereo accidental trigger rate (shown as the crossing point between the two red lines). What we aim to do with the algorithm we are presenting in this work is to reduce the accidental stereo trigger rate to 10\% of its value (open circles). We would then move to operate to the crossing point between the extrapolation of the stereo cosmic ray rate and the 10\% of the accidental stereo trigger rate (the crossing point between the two blue lines).

\end{minipage}

\section{The Topo-trigger}
\label{topo_trigger_concept}

The trigger logic implemented at this moment in the MAGIC telescope discriminate between showers and Night Sky Background (\acs{NSB}) using the spatial and time information of the pixels' signals for the single-telescope trigger level and the time information at the stereo level. To get rid of additional accidental triggers, we could also use the spatial information at the stereo level. 


\subsection{Setup of \acs{MC} simulations}
\label{simulations}
For the simulation of gamma rays we used \acs{CORSIKA} \cite{corsika} software. The particles simulated are $\gamma$-ray photons with energies ranging between 10 GeV and 30 TeV, simulated with a power-law function with 1.6 photon spectral index. For all the calculations, the spectrum was re-weighted to a Crab-like spectrum with photon spectral index $\Gamma$=2.6. The events are simulated at a $0.4^\circ$ distance from the center of the camera, as it is the standard in MAGIC observations. The \acs{Zd} ranges from 5 to 35 degrees, the \gls{Az} angle ranges from 0 to 360 degrees and the maximum impact parameter simulated is 350 meters. We used 3$\times$10$^{6}$ showers.

The current \acs{DT} applied in the \acs{L0} individual pixel trigger currently used for the MAGIC simulations ({\it Nominal} DT) is 4.5 phe for MAGIC~I (M~I) and 4.7 phe for MAGIC~II (M~II). The different \glsplural{DT} used for the two telescopes are due to the differences in reflectivity of the mirrors and \gls{QE} of the PMTs. The \acs{L1} trigger logic is called 3\acs{NN} and the \acs{L3} gate used is 180 ns. The \glsplural{DT} used to test the Topo-trigger are chosen such that if we manage to reduce the stereo trigger rate due to accidentals one order of magnitude, the stereo trigger rate is the same as for the Nominal \acs{DT}. We reduced the \acs{DT} to 4.2 phe in M~I and 4.3 phe in M~II ({\it Reduced} DT). The results for the stereo accidental trigger rates for {\it Reduced} and {\it Nominal} \glsplural{DT} can be found on Table \ref{tab:L1_rates}. The differences between the accidental rates of M~I and M~II lies on the better reflectivity of M~II mirrors and is also reproduced in the data. To partially compensate this difference, M~II \acp{DT} are $\sim$5\% higher than M~I ones.


\begin{table}[htb]
\centering
\begin{tabular}{c|c|c|c}

\hline\hline
\multicolumn{1}{c|}{} & \multicolumn{3}{|c}{Accidental trigger rate [kHz]}\\ \hline
 DT   & M~I  & M~II  & Stereo\\ \hline

Nominal & 25 $\pm$ 4 & 39 $\pm$ 5  & 0.18 $\pm$ 0.04\\ \hline

Reduced  & 78 $\pm$ 7 & 125 $\pm$9 & 1.8$\pm$ 0.2\\ 

\end{tabular}
\caption[\acs{L1} trigger rates for different \acs{NSB}, \glsplural{DT} and for the two MAGIC telescopes.]{\acs{L1} trigger rates for different \acs{NSB}, \glsplural{DT} and for the two MAGIC telescopes. The Nominal \acs{DT} corresponds to 4.5 phe for M~I and 4.7 phe for M~II, while the Reduced DT corresponds to 4.2 phe for M~I and 4.3 phe for M~II. }
\label{tab:L1_rates}
\end{table}


\subsection{Spatial information available at trigger level}

The basic idea is to implement online cuts on the spatial information available at the trigger level. In particular, we can use the 19 \acs{L1} trigger macrocell bits from the two telescopes to obtain information about the location of the image in the camera. 
When a \acs{L1} trigger is issued in each telescope, a copy of the signal goes to the prescaler and another one to the \acs{L3} trigger. We intend to deliver another copy to the Topo-trigger. The Topo-trigger compares the macrocells triggered in each telescope every time a \acs{L1} trigger is issued and sends a veto signal to the prescaler when the combination of macrocells does not correspond to that triggered by a gamma ray.



    \begin{figure}[t]
  \centering
  \includegraphics[width=0.8\textwidth]{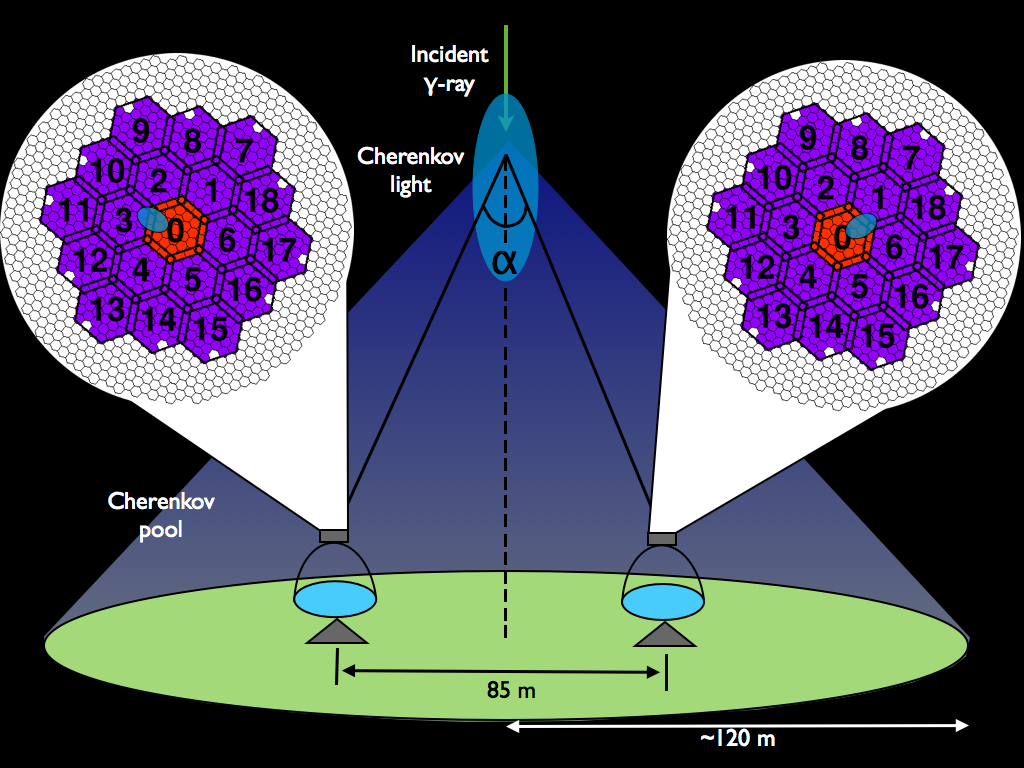}
  \caption[Scheme of the detection of Cherenkov light produced by a low energy $\gamma$-ray shower by the MAGIC telescopes.]{Scheme of the detection of Cherenkov light produced by a low energy $\gamma$-ray shower by the MAGIC telescopes. The angle $\alpha$ between the light arriving to both telescopes is smaller than $0.6^\circ$ for all the showers produced at a height of $\sim$ 10 km.}
  \label{Shower_macrocells}
 \end{figure}

\subsection{Macrocell selection}
\label{macrocell_selection}
Let us discuss the separation angle under which the shower is seen from the two MAGIC telescopes. We denote this angle alpha as is shown in Figure \ref{Shower_macrocells}. The angle is maximum when the shower develops in between the two telescopes. The distance between M~I and M~II is 85 m and the assumed distance at which the shower is produced is 10 km \acs{a.s.l.} ($\sim$7.8 km above the telescopes). The maximum angle separation of a point-like shower between the two telescopes is $\alpha\simeq0.6^\circ$. Since every pixel covers 0.1$^\circ$, we conclude that the maximum separation of the showers seen in the two MAGIC cameras is 1 macrocell. We have to point out that the angle $\alpha$ calculated depends on the height at which the shower interacts with the atmosphere (high energy showers go deeper into the atmosphere, therefore produce larger $\alpha$ angles). This means that most of the showers trigger the same macrocell at both telescopes or neighboring macrocells. This is the essence of our new trigger level.

\begin{figure*} [!htb]   
\begin{minipage}[]{0.5\textwidth}
\includegraphics[width=0.9\linewidth]{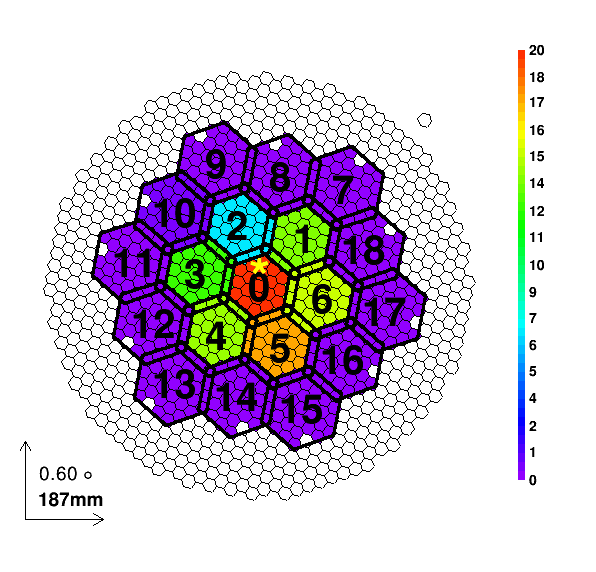}
\end{minipage}
\hspace{\fill}
\begin{minipage}[]{0.5\textwidth}
\includegraphics[width=0.9\linewidth]{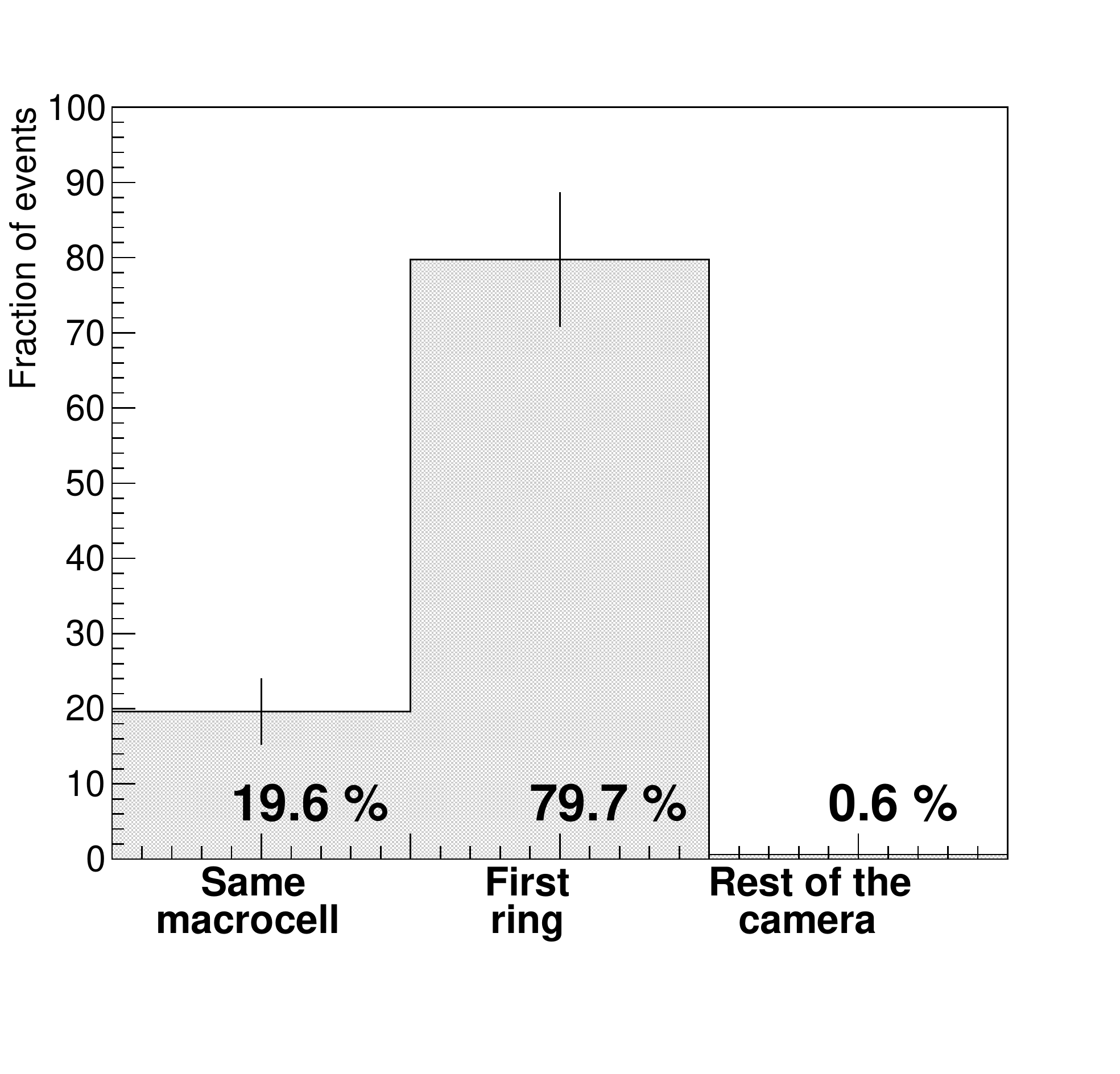}
\end{minipage}

 \caption[Macrocells triggered in M~II when selecting one macrocell in M~I for events triggering only one macrocell in each telescope.]{Macrocells triggered in M~II when selecting one macrocell in M~I for events triggering only one macrocell in each telescope for simulated gamma rays after analysis cuts (left panel) and the histograms of the distributions (right panel).}
 \label{Macrocells_and_histograms}
\end{figure*}

In the simulations, we record the macrocell digital output for 10 ns after the \acs{L1} trigger is issued. This digital output of each macrocell is 0 if the macrocell was not triggered during those 10 ns and 1 if it was triggered. As the events triggered by \acs{NSB} are accidentals, we expect them to trigger only one macrocell. Using \acs{MC} simulations, we calculated that  the probability that an accidental event is triggered by more than one macrocell is $\textrm{P}_{2\textrm{M}}=0.4\%\cdot\textrm{P}_{1\textrm{M}}$, where $\textrm{P}_{2\textrm{M}}$ is the probability of triggering 2 macrocells due to an accidental and $\textrm{P}_{1\textrm{M}}$ the probability of triggering 1. As the fraction of events triggering more than one macrocell is much smaller than the one triggering only one, we selected the events that triggered only one macrocell in each telescope and studied them. In order to illustrate possible macrocell 1--1 combinations for gamma rays, we select showers that gave triggers in a given macrocell in M~I and look at the macrocell distribution for these showers in M~II. In Figure \ref{Macrocells_and_histograms} we have two examples: in the top panel we have selected events for which only macrocell 0 (the central one, marked with an asterisk) is triggered in M~I. On the bottom panel, we have selected events for which only macrocell 17 (one of the border ones) is triggered. The color of the left panel plots represents the fraction of events that triggered a given macrocell in M~II. In the right panel plots we show how the events distribute in M~II. Events in the first bin triggered the same macrocell in M~I and M~II (macrocell 0). Events in the second bin triggered in M~II one of the macrocells of the first ring surrounding macrocell 0, in this case, macrocells 1, 2, 3, 4, 5 and 6. The third bin corresponds to the rest of the macrocells.

From the events triggering only one macrocell in each telescope, we will accept only events triggering the same macrocell in both telescopes or the neighboring ones. We will keep most of the events triggered by gamma rays, while getting rid of a large fraction of the accidental stereo triggers.
Each telescope has N=19 macrocells, so there are N$^2$=361 possible different combinations of one-to-one macrocell. The Topo-trigger, in first approximation, accepts 103 of them. Showers are mainly distributed according to the distribution we have obtained with the simulations, but the triggers due to accidentals will be randomly distributed in the whole camera, and most of them will be triggering only one macrocell each time. 


We will apply further cuts based on the position of the macrocells respect to the source.
Following the geometry shown in Figure \ref{Shower_macrocells}, if the source is at a certain direction in the sky, we do not only know that the shower should fall in the same macrocell or the surrounding ones, but also which of the surrounding macrocells will be hit in the other telescope. If we divide the data in bins of  \acs{Az}, we can further select the macrocells that are accepted and increase our rejection power. We could make Az bins as fine as desired, but at some point there is no improvement. As the finer possible binning would be given by the size of the macrocells, since we have 12 outer macrocells  we cannot improve further than establishing 12 bins in \acs{Az}. By applying the macrocell selection to each \acs{Az} bin, we obtain:

$$\genfrac{}{}{0pt}{}{\mathrm{\%\ Fraction\ of}} {\mathrm{\ rejected\ accidentals}}=\frac{361-53}{361}\times100=85\%$$

Let us now evaluate the impact of the cuts on the \g-ray events.
We have applied the Topo-trigger macrocell selection to \acs{MC} \g-ray events triggering with the Reduced \acs{DT} configuration. The fraction of simulated gamma rays rejected by the algorithm is $\sim$ 2.4 \%. As we will show in Section \ref{performance_topo}, the fraction of events rejected at the analysis level is significantly lower than this. In summary, according to \acs{MC}, by applying the Topo-trigger macrocell selection cuts, we reject 85\% of the accidental events but only 2.4 \% of the \g-ray events at the trigger level.


\subsection{Expected performance}
\label{performance_topo}
We will now calculate the collection area and energy threshold of the instrument after applying the Topo-trigger selection. For the analysis of the \gls{MC} gamma rays, we calibrate the data, clean the images applying the image cleaning method and apply a gamma/hadron separation with the so-called random forest algorithm as in the standard MAGIC analysis \cite{MARS}. We ran \gls{MC} simulations for 2 cases, and for the second case we apply two different image cleanings:

\begin{enumerate}[a)]

\item Nominal \acs{DT} with standard image cleaning. 

\end{enumerate}

\begin{enumerate}[b.1)]

\item Reduced \acs{DT} with the standard image cleaning.

\item Reduced \acs{DT} with an image cleaning with charge parameters reduced by 7 \%.

\end{enumerate}

\begin{minipage}[h]{0.48\textwidth}
  \includegraphics[width=\textwidth]{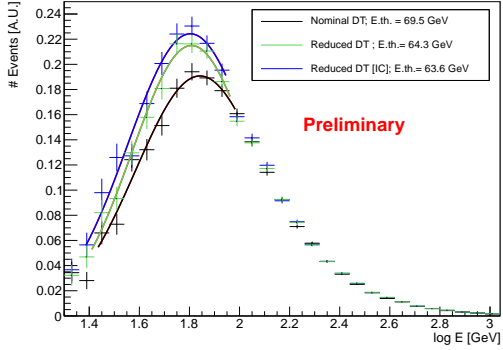}
\captionof{figure}{True energy distribution of \acs{MC} gamma rays (in arbitrary units) for a source with a 2.6 spectral index for different trigger configurations.}
\label{Energy_threshold_Turku}
\vspace{0.5 cm}
\end{minipage}
 \hspace{0.01\textwidth}
\begin{minipage}[h]{0.48\textwidth}
Figure \ref{Energy_threshold_Turku} shows the true energy distribution of \acs{MC} \g-ray events for the different trigger configurations. The black histogram represents the rate for the current trigger configuration of MAGIC (Nominal \acs{DT}). The green histogram represents the rate for the Reduced \acs{DT} configuration with \acs{DT}=4.2 phe for M~I and \acs{DT}=4.3 phe for M~II applying Topo-trigger macrocell selection with the standard image cleaning. The blue line represents the rate for the Reduced \acs{DT} configuration with a 7\% reduced image cleaning. We can see that the energy threshold goes down by up to 8 \% at the analysis level.

\end{minipage}

\begin{minipage}[h]{0.48\textwidth}
The collection area for both the Nominal \acs{DT} and the Reduced DT is shown in the top panel of Figure \ref{collection_Turku}, and  the ratio between the collection area obtained using the current MAGIC trigger configuration and the collection area obtained with the Reduced \acs{DT} applying the Topo-trigger macrocell selection in the bottom panel of the same figure. The black points correspond to the collection area obtained with the current trigger configuration of MAGIC. The green points correspond to the collection area obtained reducing the \acs{DT} to 4.2 phe in M~I and 4.3 phe in M~II and applying the Topo-trigger macrocell selection. The blue points represent the rate for the Reduced \acs{DT} configuration with a 7\% reduced image cleaning. The red line in the bottom panel delimits the region where the Reduced \acs{DT} option performs better than the Nominal \acs{DT} (ratio $>$ 1). 
\end{minipage}
 \hspace{0.01\textwidth}
\begin{minipage}[h]{0.48\textwidth}
  \includegraphics[width=\textwidth]{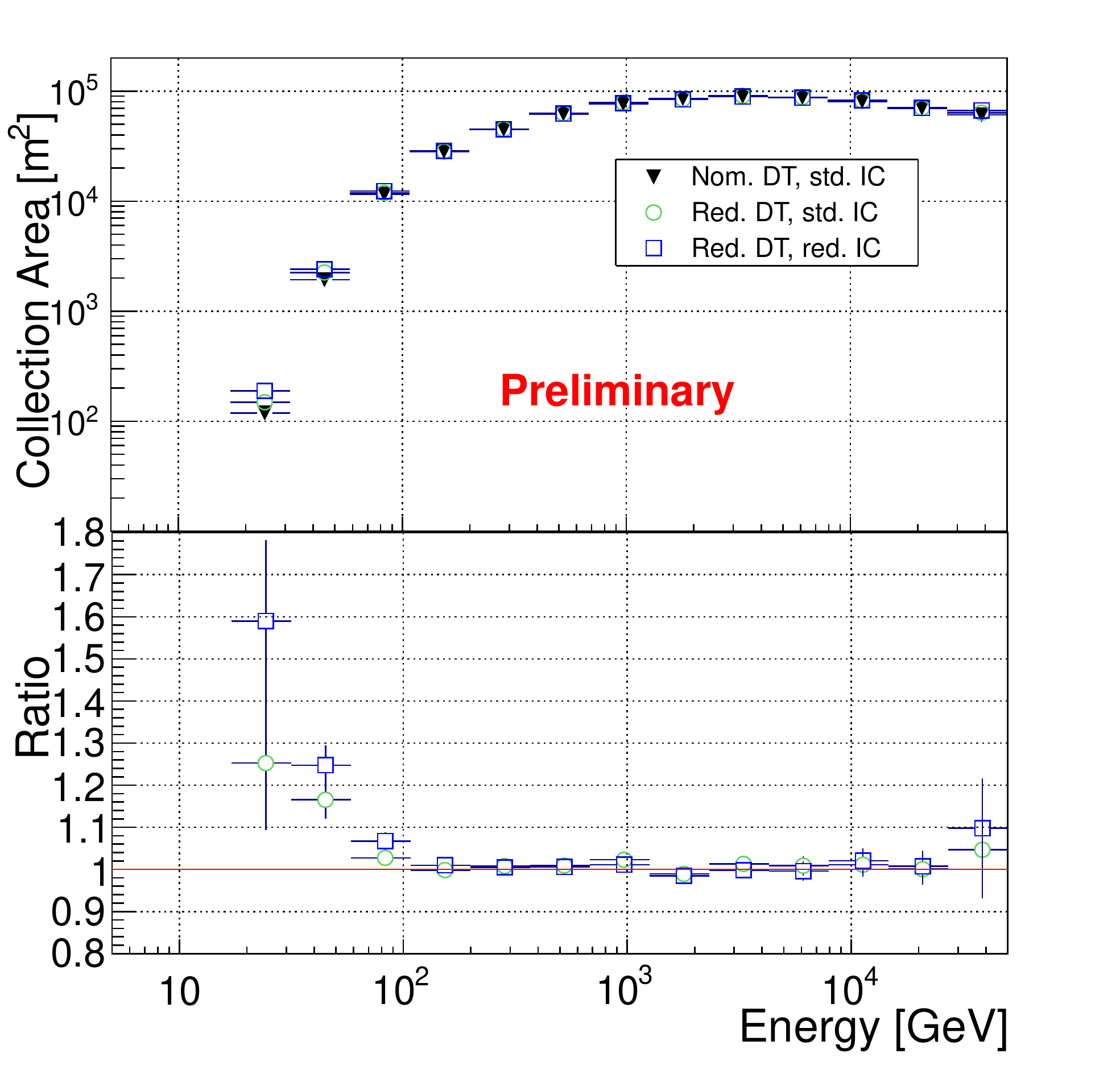}
\captionof{figure}{Collection area using the Nominal \acs{DT} and the Reduced DT applying cuts (top) and the ratio between the Reduced \acs{DT} configurations and the Nominal \acs{DT} one (bottom).}
\label{collection_Turku}
\vspace{0.5 cm}

\end{minipage}

We can see that the improvement in collection area using the Topo-trigger  at the lowest energies is $\sim$60\%, although the errors are very large due to the low statistics at those energies. At the energy threshold, where we have a peak in the number of events triggered, the improvement using the Topo-trigger is between 10--20 \% with respect to the current MAGIC configuration. Moreover, this gain in events is still $\sim$5\% for showers with energies between 70 GeV to 100 GeV. At the highest energies (the last point in energy of Figure \ref{collection_Turku}), there are only a few events that give trigger, therefore the fluctuations are large. We would like to stress that all these improvements are at the analysis level, i.e. for the events that are used to derive spectra, light curves and skymaps. After applying the Topo-trigger cuts at the analysis level, we keep 98.8\% of the total events that survived the analysis cuts.

\section{Piggy-back measurements}
\label{telescope_data}

We installed a system to record the macrocell pattern for every event at the trigger system at the MAGIC telescopes. As the trigger \acs{DT} was not lowered, we do not expect any improvement by applying the algorithm to the data. We only intend to verify that no gamma rays are lost at the analysis level due to the Topo-trigger macrocell selection applied in the software and hardware.




We took Crab Nebula data at \acs{Az} angles ranging from 100$^\circ$ to 175$^\circ$ and \acs{Zd} between 6$^\circ$ and 22$^\circ$. We analyzed these data using the standard analysis in MAGIC and applied the standard hadron/gamma separation and event reconstruction cuts for two energy ranges: medium-to-high energies and low energies. 

\paragraph{Medium-to-high energies}

We applied the standard $\gamma$/hadron separation, reconstructed energy and $\theta^2$ cuts used for medium energies, leading to an energy threshold of $\sim$ 250 GeV. We calculated the sensitivity of the telescope and the $\gamma$-ray rate for the sample without applying any cut in macrocells and applying the Topo-trigger macrocell cuts. The result was that we have exactly the same number of background and gamma events after applying the Topo-trigger macrocell selection cuts, so the sensitivity and \g-ray rate are kept constant.






\paragraph{Low energies}

We also made an analysis applying the standard cuts for low energies.  We computed the sensitivity ratio (Sensitivity [macrocell\ cuts]/Sensitivity [no\ cuts]) and \g-ray rate ratio (\g-ray rate [macrocell\ cuts]/\g-ray rate [no\ cuts]) with and without macrocell cuts. The results can be found in Table \ref{Crab_summary}. They confirm what we expected from the simulations: if we apply the macrocell selection to the events used for analysis, we keep basically the same detection efficiency.

\begin{table}[h!tb]
\begin{center}
\begin{tabular}{c|c|c}
& Sensitivity ratio & $\gamma$-ray rate ratio\\
 \hline
 \hline
 Medium E   & 1  &  1  \\
 \hline
Low E  & 1.005 $\pm$ 0.007 & 0.997 $\pm$ 0.001 \\
\end{tabular}
\end{center}
\caption{Summary of the results for the sensitivity ratio and $\gamma$-ray rate ratio between the analysis applying the Topo-trigger macrocell cuts and without applying them for medium and low energies.}
\label{Crab_summary}
 \centering
\end {table}

\section{Discussion and conclusions}
\label{discussion}

We developed a novel stereo trigger system for \acp{IACT} which make use of the topological information of the showers in the camera. Combining the information of the \acs{Az} angle at which the telescope is pointing and the \acs{L1} trigger macrocell hit in each telescope we can reject 85\% of the accidental stereo trigger rate, which is the dominant at the lowest energies, without losing gamma rays. By studying the effect of applying the selection algorithm to off-axis data, we find that the discrimination power of the algorithm does not depend on the source position in the camera. We run simulations reducing the \acs{DT} used for triggering the telescopes and applying this algorithm and we found that implementing this trigger translates into a decrease of up to 8\% in the energy threshold and an increment of  $\sim$60\% in the collection area at the lowest energies and from 10-20\% at the energy threshold, where most of the events are triggered. We installed a device to record the triggered macrocells of the events recorded by the MAGIC telescope. Without reducing the \acs{DT} applied at the \acs{L0} trigger, we verified that the Topo-trigger macrocell selection tested in the \acs{MC} does not lead to any loss in the sensitivity or in the $\gamma$-ray rate. 

The board that will be used to veto signals from the \acs{L3} trigger is already installed in the MAGIC telescope and is currently under commissioning.

\begin{acknowledgments}

The authors would like to thank the support of the MAGIC collaboration. We would also like to thank the Instituto de Astrof\'{\i}sica de Canarias for the excellent working conditions at the Observatorio del Roque de los Muchachos in La Palma. This work is partially funded by the ERDF under the Spanish MINECO grant FPA2012-39502.
\end{acknowledgments}

\bibliographystyle{JHEP}
\bibliography{references} 


\end{document}